%% file: main.tex
\documentclass[sigconf]{acmart}

\usepackage[utf8]{inputenc}

\usepackage{latexsym}
\usepackage{wasysym}

\usepackage{balance}

\usepackage{booktabs} 
\usepackage{xspace}
\usepackage{url}
\usepackage{enumerate}
\usepackage{multirow}
\usepackage{array}
\usepackage{lscape}
\usepackage{adjustbox}
\usepackage{amssymb,amsmath,amsthm}

\usepackage{soul}

\input{macros}

\begin{document}

\copyrightyear{2018}
\acmYear{2018}
\setcopyright{acmlicensed}
\acmConference[SIGIR '18]{The 41st International ACM SIGIR
Conference on Research & Development in Information Retrieval}{July
8--12, 2018}{Ann Arbor, MI, USA}
\acmBooktitle{SIGIR '18: The 41st International ACM SIGIR Conference
on Research \& Development in Information Retrieval, July 8--12, 2018,
Ann Arbor, MI, USA}
\acmPrice{15.00}
\acmDOI{10.1145/3209978.3210024}
\acmISBN{978-1-4503-5657-2/18/07}
\fancyhead{}

\title{An Axiomatic Analysis of Diversity Evaluation Metrics: Introducing the Rank-Biased Utility Metric}

\author{Enrique Amig{\'o}}
\affiliation{
  \institution{NLP \& IR Group at UNED}
  \city{Madrid}
  \country{Spain}
}
\email{enrique@lsi.uned.es}

\author{Damiano Spina}
\affiliation{
  \institution{RMIT University}
  \city{Melbourne}
  \country{Australia}
}
\email{damiano.spina@rmit.edu.au}

\author{Jorge Carrillo-de-Albornoz}
\affiliation{
  \institution{NLP \& IR Group at UNED}
  \city{Madrid}
  \country{Spain}
}
\email{jcalbornoz@lsi.uned.es}

\begin{abstract}
Many evaluation metrics have been defined to evaluate the effectiveness ad-hoc retrieval and search result diversification systems. However, it is often unclear which evaluation metric should be used to analyze the performance of retrieval systems given a specific task. Axiomatic analysis is an informative mechanism to understand the fundamentals of metrics and their suitability for particular scenarios. 
 In this paper, we define a  constraint-based axiomatic framework  to study the suitability of existing metrics in search result diversification scenarios. 
The analysis informed the definition of \emph{Rank-Biased Utility (RBU)} -- an adaptation of the well-known Rank-Biased Precision metric -- that takes into account redundancy and the user effort associated to the inspection of documents in the ranking. Our experiments over standard diversity evaluation campaigns show that the proposed metric captures quality criteria reflected by different metrics,  being suitable in the absence of knowledge about particular features of the scenario under study.
\end{abstract}



\maketitle

\input{01_introduction}
\input{02_background}
\input{03_formal_cons}
\input{04_analysis}

\input{05_proposal}
\input{06_experiments}
\input{07_conclusions}

\myparagraph{Acknowledgments}
This research was partially supported by the Spanish Government (project Vemodalen TIN2015-71785-R) and the Australian Research Council (project LP150100252). The authors wish to thank the reviewers for their valuable feedback.

\bibliographystyle{ACM-Reference-Format}
\bibliography{strings-shrt,diversity_2018} 
\appendix
\input{08_Appendix}
\end{document}

%% file: macros.tex
\newcommand{\tick}{\CIRCLE}
\newcommand{\fail}{\Circle}

\newtheorem{cons}{Constraint}

\newcommand{\rand}[0]{\text{\it rand}}

\newcommand{\Pri}[0]{\textsc{Pri}\xspace}
\newcommand{\Deep}[0]{\textsc{Deep}\xspace}
\newcommand{\DeepTh}[0]{\textsc{DeepTh}\xspace}
\newcommand{\CloseTh}[0]{\textsc{CloseTh}\xspace}
\newcommand{\Conf}[0]{\textsc{Conf}\xspace}

\newcommand{\Nov}[0]{\textsc{AspDiv}\xspace}
\newcommand{\Red}[0]{\textsc{Red}\xspace}
\newcommand{\MRed}[0]{\textsc{MRed}\xspace}
\newcommand{\Sat}[0]{\textsc{Sat}\xspace}
\newcommand{\AspRel}[0]{\textsc{AspRel}\xspace}

\newcommand{\MU}[0]{$\mbox{MU}$\xspace}

\newcommand{\myparagraph}[1]%
  {\paragraph*{\hspace*{-\parindent}\normalsize\bf#1.}}
\newcommand{\mycaption}[1]{\caption{{\rm{#1}}}}



%% file: 01_introduction.tex
\section{Introduction}
\label{sec:introduction}

The development of better information retrieval systems is driven by how improvements are measured. The design of test collections and evaluation metrics that started with the Cranfield paradigm in the early 1960s allowed researchers to analyze the quality of different retrieval models in an automated and cost-effective way. Since then,  many  
evaluation metrics have been proposed to measure the \emph{effectiveness} of information retrieval systems~\cite{sparck1976information, voorhees2005trec, scholer2016information}.

Selecting a suitable set of metrics for a specific task is challenging. 
Comparing metrics empirically against user satisfaction or search effectiveness 
requires data that is often unavailable. Moreover, findings may be biased to the subjects, retrieval systems or other experimental factors. 

An alternative consists of modeling theoretically the desirable properties of retrieval systems, as well as the abstraction of the expected users' behavior when performing a specific task. For instance, 
a metric that looks at how early the relevant document is retrieved in the ranking -- such as Reciprocal Rank~\cite{Voorhees99thetrec-8} --  would be an appropriate metric to analyze the performance of systems on a single-item navigational task.
However, is often challenging to come up with the proper evaluation tools for more complex search scenarios, as is the case of search result diversification~\cite{santos2015search}. In this context, the ranking of retrieved documents must be optimized in such a way that diverse query aspects are captured in the first positions. The challenge is that the evaluation of system outputs is affected by multiple variables such as: the deepness of ranking positions, the amount of documents in the ranking related to the same query aspect, relevance grades, the diversity of query aspects captured by single documents or the user's effort when inspecting the ranking.

Axiomatic analysis has been shown to be an effective methodology to better understand the foundamentals of evaluation metrics~\cite{van1974foundation,busin2013axiometrics,Amigo_2013, Ferrante:2015}. In the context of evaluation, axiomatic  approaches consist of a verifiable set of formal constraints  that reflect which quality factors are captured by metrics, facilitating the metric selection in specific scenarios.
To our knowledge, there is no comprehensive axiomatic analysis of the behavior of diversity metrics in the literature. This paper provides a set of ten formal constraints that focus on both retrieval and diversity quality dimensions.  

We found that every constraint is satisfied at least by one metric. However, none of the existing diversity metrics satisfy all the proposed constraints simultaneously. 
In order to solve this gap, we define the metric \emph{Rank-Biased Utility (RBU)} by integrating components from different metrics in order to capture every formal constraints. RBU is an adaptation of the well-known Rank-Biased Precision metric~\cite{RBP} that incorporates redundancy and the user's effort associated to the inspection of documents in the ranking. Our experiments using standard diversity test collections validate our axiomatic analysis.  Results show that, 
satisfying every constraint with a single metric leads to \textit{unanimous} evaluation decisions when compared against other existing metrics,  
i.e., RBU captures quality criteria which are reflected by different metrics. Therefore, this metric offers a solution in the absence of knowledge about the specific characteristic of a diversity-oriented retrieval scenario. Moreover, the theoretical framework presented in this paper helps to decide which metric should be used.

The paper is organized as follows. Section~\ref{sec:relatedwork} describes related work on evaluation of evaluation metrics. Section~\ref{sec:constraints} introduces the formal constraints that we propose to analyze relevance and diversity properties of metrics. Section~\ref{sec:metric:analysis} provides a comprehensive analysis of existing diversity metrics according to these constraints and Section~\ref{sec:rbu} defines the proposed RBU metric. Section~\ref{sec:metaevaluation} details the results of our experiments. Finally, Section~\ref{sec:conclusions} concludes the work.

%% file: 02_background.tex
\section{Related Work}
\label{sec:relatedwork}

There is no consensus of meta-evaluation criteria for search result diversification. Some  works  inherit meta-evaluation criteria from  ad-hoc metrics such as \emph{sensitivity} to system differences~\cite{sakai2011evaluating, Luo-13, Sakai-10, Golbus-13}. This methodology however does not give information about to what extent metrics capture diversity properties.  {\citet{Smucker-12}} studied the correspondence between metric scores and user effort when exploring document rankings. This methodology has the advantage of being realistic -- effort is calibrated from historical log data -- but only focuses on partial quality aspects.

Most of works on diversity metrics are supported by descriptive analysis. In 2008, {\citet{Clarke-08}} meta-evaluated $\alpha$-nDCG  by analyzing the effect of modifying the diversity parameter $\alpha$ under different datasets. 
One year later, {\citet{Agrawal-09}} checked the \emph{intent-aware} scheme for diversification by studying the evaluation results of three search engines.
{\citet{Clarke-09}} proposed Novelty- and Rank-Biased Precision (NRBP), an extension of RBP~\cite{RBP} for diversification, 
joining properties of the original RBP metric, $\alpha$-nDCG and intent-aware metrics. 
In 2010,  {\citet{Sakai-10}} compared their proposed approach to $\alpha$-NDCG and NRBP, in terms of metric agreement under different 
parameters.  The authors considered some meta-evaluation criteria such as interpretability, computability or capability to 
accommodate graded relevance and score ranges.
Three years later,  \citet{chandar2013preference} evaluated their approach by studying correlation with previous metrics while reflecting other ranking quality issues.
 {\citet{Luo-13}} proposed the Cube Test metric. They  studied the effect
of the metric parameters under synthetic system outputs, in the same manner than {\citet{Clarke-08}}. \citet{Tangsomboon-14} in 2014 and also \citet{Yu-17} in 2017,  supported their proposed metrics in terms of  agreement and disagreement with previous metrics. 

Not many works define a way of quantifying the suitability of metrics to capture diversity. An exception is the work by \citet{Golbus-13}
who defined \emph{Document Selection Sensitivity}. This meta-measure
 reports to what extent metrics  are sensitive to document rankings containing  relevant documents  but different grades of diversity. Within this line, we define in this work \emph{Metric Unanimity} (\MU), which quantifies to what extent a metric is sensitive to quality aspects captured by other existing metrics.

On the other hand, metrics have been successfully analyzed in terms of formal constraints  in ad-hoc retrieval scenarios 
\cite{Amigo_2013, moffat:2013, Ferrante:2015}. The axiomatic  methodology 
consists of identifying theoretical situations in which metrics should behave in a particular manner. This methodology 
has several strengths: it is objective,  independent from
datasets and it facilitates the interpretation of metrics. We found only a few initial works in the context of formal constraints for search result diversification. For instance,  {\citet{Leelanupab-13}} reviewed the appropriateness  of intent-aware, stating an extreme particular situation in which ERR-IA does not behave as expected. In our work, we meta-evaluate existing metrics  on the basis of ten  constraints that formalize desirable properties for ranking and diversity effectiveness.

%% file: 03_formal_cons.tex
\section{Axiomatic Constraints}
\label{sec:constraints}

\subsection{Problem Formalization}

We formalize the output of a document retrieval system 
as an ordered list of documents $\vec{d}=(d_1,\ldots,d_n)$ of length $n$, extracted from a collection of documents $\mathcal{D}$.  In order to express formal constraints, 
we use $\vec{d}_{i \leftrightarrow j}$ to denote the result of 
swapping documents between positions $i$ and $j$. Likewise, 
$\vec{d}_{d \leftrightarrow d'}$ denotes the result of replacing the document $d$
with the document $d'$ in the ranking $\vec{d}$.

For search result diversification, we consider a set of query aspects 
$\mathcal{T}=\{t_1,\ldots,t_m\}$. For instance,  users searching for a restaurant may be interested in the menu, the offers, opening times, etc.  Each aspect has an associated \emph{weight} $w(t_j)$ and 
the sum of all aspect weights adds up to $1$: $\sum_{j=1}^m w(t_j)=1$.

On the other hand, $r(d_i,t_j)\in[0\ldots 1]$ represents the graded \textit{relevance} of document $d_i$ to the aspect $t_j$.  We assume the user's behavior follows the \emph{cascade model}, i.e., the user inspects the ranking sequentially from the top to the bottom, until either (i) the user's information needs get satisfied or (ii) the user stops looking (i.e., user's patience is exhausted). Following the same user model than the one used by Expected Reciprocal Rank~\cite{chapelle2009expected}, we consider \textit{relevance} as the suitability of the document to satisfy the user needs, which has a negative correspondence with the probability of exploring more documents. Finally, we use $Q(\vec{d})$ to denote the ranking quality score, i.e., the score given by applying an evaluation metric $Q$ to a given ranking $\vec{d}$.

Our axiomatic approach consists of a set of ten
formal constraints that evaluation metrics may satisfy. These constraints are grouped into two sets: \emph{relevance-oriented} and \emph{diversity-oriented}, that we describe below.

In the definition of the constraints, we may refer to the following conditions: 
\emph{single aspect} $(|\mathcal{T}|=1)$; 
\emph{balanced aspects} $(\forall t \in \mathcal{T} \ldotp w(t)=1/|\mathcal{T}|)$;
\emph{binary relevance} $(\forall t,d \ldotp r(d,t) \in \{0,r_c\})$;
\emph{no aspect overlap} $(r(d,t) > 0 \Rightarrow \forall t' \neq t \ldotp r(d,t') = 0)$;
 and \emph{relevance contribution} $\left( r \left( d, t \right) \ll 1 \right)$.  The last condition means that finding new relevant documents about the same topic is always effective. In other words, there is always room for new documents to fully satisfy the user needs.

\subsection{Relevance-Oriented  Constraints}

In order to isolate relevance from diversity and redundancy, for these constraints we will assume 
\emph{single aspect}
and \emph{relevance contribution}. 

For the sake of legibility, we use the notation:
$r(d)=r(d,t)$. We also denote $d^{\mbox{\it rel}}$ and $d^{\neg
  \mbox{\it rel}}$ as relevant and non-relevant documents, respectively. That is:
$\forall i\in 1..n \ldotp r(d_i^{\neg rel})=0$ and $r(d_i^{rel})=r_{c}$. 
Under these assumptions, we import the five
constraints proposed by {\citet{Amigo_2013}} which capture previous axiomatic 
properties~\cite{moffat:2013, Ferrante:2015}. 

\begin{cons}
[Priority, Pri] Swapping items in concordance with
their relevance increases the ranking quality score. Being  $k>0$:
\begin{equation}\label{eq:Pri}
r\left(d_{i+k}\right)>r\left(d_i\right)\Longrightarrow Q\left(\vec{d}_{i\leftrightarrow i+k}\right)>Q\left(\vec{d}\right)
\end{equation}
\end{cons}

The next constraint is based on the intuition that the
effect of relevance  depends on the document ranking position.  
This constraint is also referred as \textit{top-heaviness}:

\begin{cons}
[Deepness, Deep]  Correctly swapping contiguous items has more effect
in early ranking positions:
\begin{equation}\label{eq:Deep}
\resizebox{.9\hsize}{!}{%
$r(d_i)=r(d_j) < r(d_{i+1})=r(d_{j+1}) \Longrightarrow 
Q\left(\vec{d}_{i\leftrightarrow i+1}\right) > Q\left(\vec{d}_{j\leftrightarrow j+1}\right)$
}
\end{equation}
where $i<j$.
\end{cons}

The next constraint reflects that the effort spent by the user to inspect 
a long (deep) list of search results is limited. In other words, there is an 
area of the ranking that  may never get explored by the user:
\begin{cons}
[Deepness Threshold, DeepTh] Assuming binary relevance, there
exists a value $n$ large enough such that, retrieving  only one relevant document
at the top of the ranking is better than retrieving $n$ relevant documents
after $n$ non-relevant documents:
\begin{equation}\label{eq:DeepTh}
\exists n\in\mathbb{N}^+ \ldotp Q\left(d_1^{\mbox{rel}},\ldots\right) > Q\left(d_1^{\neg \mbox{rel}}, \ldots , d_n^{\neg \mbox{rel}} , d_1^{ \mbox{rel}}, \ldots , d_n^{ \mbox{rel}}\right)
\end{equation}
\end{cons}

On the other hand, we can assume that there exists a (short) ranking area which is 
always explored by the user. In other words, at least a few  documents are inspected by the user 
with a minimum effort. This means that, at the top of the ranking, the amount of
captured relevant documents is more important than their relative rank positions.

\begin{cons}
[Closeness Threshold, CloseTh] Assuming binary relevance, there
exists a value $m$ small enough such that retrieving one relevant document
in the first position is worse than $m$ relevant documents after $m$ non-relevant documents: 
\begin{equation}\label{eq:CloseTh}
\exists m\in\mathbb{N}^+ \ldotp Q\left(d_1^{\mbox{rel}}, \ldots\right) < Q \left(d_1^{\neg \mbox{rel}}, \ldots  , d_m^{\neg \mbox{rel}}, d_1^{\mbox{rel}}, \ldots , d_m^{\mbox{rel}} \right)
\end{equation}
\end{cons}

In some particular scenarios, however, this may not hold. For instance, in audio-only search scenarios, search results may be delivered sequentially one-at-a-time.

Finally, the amount of documents returned is also an aspect of the system quality. 
In the same manner that capturing diversity in the first positions is desirable, 
adding non-relevant documents to the end of the ranking should be penalized by metrics.
In other words, the cutoff used by the system to \emph{stop} returning search results has also an impact on users. Therefore, adding noise at the bottom of the ranking should decrease its effectiveness.

\begin{cons}
[Confidence, Conf] Adding non-relevant documents decreases the 
score:
\begin{equation}\label{eq:Conf}
Q\left(\vec{d}\right) > Q\left(\vec{d}, d^{\neg \mbox{rel}}\right)
\end{equation}
\end{cons}

\subsection{Diversity-Oriented Constraints}


The first diversity-oriented constraint is related to the fact that the metric should be sensitive
to the \emph{novelty} of aspects covered by a single document:

\begin{cons}
[Query Aspect Diversity, \Nov] Covering more aspects in the same document (i.e., without additional effort of inspecting more documents) increases the score. Assuming relevance contribution $(\forall d,t \ldotp r(d,t)\ll 1)$:
\begin{equation}\label{eq:Nov}
\forall t\in\mathcal{T} \ldotp 
   \left(r(d_i',t)>r(d_i,t)\right) \Longrightarrow 
    Q\left(\vec{d}_{d_i \leftrightarrow d_i'}\right) > Q \left(\vec{d}\right)
\end{equation}
\end{cons}

To calculate the \emph{gain} obtained by observing a new relevant document in the ranking, most of the existing diversity metrics take into account
 the number of previously observed documents
that are related with the same aspect. 
The more an aspect has been covered earlier in the ranking, the less a new document relevant to this aspect contributes to the gain. Formally:

\begin{cons}
[Redundancy, Red]
Assuming binary relevance, balanced aspects and no aspect overlap,  and being $d$ and $d'$ documents relevant to different aspects $r(d,t)=r(d',t')=r_c$, then:
\begin{equation}\label{eq:Red} 
\begin{split}
 |\{d_i\in\vec{d} \ldotp r(d_i,t)=r_c\}|      &   >  |\{d_i\in\vec{d} \ldotp r(d_i,t')= r_c\}|  \Longrightarrow\\
                Q\left(\vec{d},d'\right)         &   >  Q\left(\vec{d},d\right)
\end{split}
\end{equation}
\end{cons}

The \Red constraint  assumes binary relevance, by counting relevant documents for each 
query aspect. In order to consider graded relevance 
in previously observed documents, we can apply the monotonicity
principle. That is, if an aspect $t$ is captured to a greater
extent than a second aspect $t'$ in every previously observed document, 
then the ranking is more redundant w.r.t. $t$ than $t'$. Formally:

\begin{cons}
[Monotonic Redundancy, MRed]
Assuming two balanced aspects ($\mathcal{T}=\{t,t'\}$), relevance contribution, and being $d$ and $d'$ documents exclusively relevant to each 
aspect,   $0<r(d,t)=r(d',t')\ll 1$  and $r(d,t')=r(d',t)=0$:
\begin{equation}\label{eq:MRed}
\forall d_i\in \vec{d} \ldotp \left(r(d_i,t)>r(d_i,t')\right)
\Longrightarrow 
Q\left(\vec{d},d'\right) > Q\left(\vec{d},d\right)
\end{equation}
\end{cons}

Intuitively, as well as the exploration capacity or patience of the user is limited, the user's information need is also finite. 
This means that there should exists a certain point on which 
a new single piece of information completely satisfies user's information needs, 
in such a way that retrieving any other  documents addressing the same query 
aspect is not beneficial. Formally:

\begin{cons}
[Aspect Relevance Saturation, Sat]
Assuming no aspect overlap, there exists a finite relevance value $r_{max}$
 large enough such that:
\begin{align}\label{eq:Sat}
\begin{split}
(r(d_{n},t) = r_{max}) ~~\wedge ~~&(r(d_{n+1},t) > 0)  \Longrightarrow \\
Q\left(\vec{d}\right)~\ge~&Q\left(\vec{d}, d_{n+1} \right)
\end{split}
\end{align}
\end{cons}

Finally, the following constraint captures the relative weight of aspects $w(t)$
w.r.t. the user's information need:

\begin{cons}
[Aspect Relevance, AspRel] Aspects with higher weights have more effect
in score of the ranking quality. Formally, assuming no aspect overlap, and being $d_i$ and $d_i'$ documents
 that are relevant to different  aspects that have not been observed before,
$\forall j<i \ldotp r(d_j,t)=r(d_j,t')=0$, and $r(d_i,t)=r(d_i',t')>0$  then:
\begin{equation}\label{eq:AspRel}
w(t) < w(t') \Longrightarrow 
Q\left( \vec{d}_{d_i\leftrightarrow d_i'} \right) > Q\left(\vec{d}\right)
\end{equation}
\end{cons}

In summary, we have defined a total of ten constraints: five relevance-oriented constraints
(\Pri, \Deep, \DeepTh, \CloseTh and \Conf), and five constraints for search result diversification (\Nov, \Red, \MRed, \Sat, and \AspRel). The next section provides an axiomatic analysis of the most popular retrieval and diversity metrics using these constraints.

%% file: 04_analysis.tex
\section{Metric Analysis}
\label{sec:metric:analysis}

In this section, we firstly analyze standard metrics designed to evaluate retrieval systems in non-diversified scenarios (i.e., single-aspect). Then we analyze the  \emph{intent-aware} family of metrics, as well as a number of popular diversity metrics.

\subsection{Standard  Metrics for Ad-hoc Retrieval}

We analyze here   metrics  that do not consider multiple aspects of a query or topic, including:
Precision at a cutoff $k$ (P@$k$),  
Reciprocal Rank (RR)~\cite{Voorhees99thetrec-8}, 
  Average Precision (AP), 
   Rank-Biased Precision (RBP)~\cite{RBP},
Expected Reciprocal Rank (ERR@$k$)~\cite{chapelle2009expected} and 
Normalized Discounted Cumulative Gain (nDCG@$k$)~\cite{DCG}.

RBP uses a parameter $p$ that defines user's \emph{patience}, modeled as the probability
of the user to inspect the next document in the ranking. P@$k$, ERR and nDCG
include a cutoff $k$ that limits the rank
positions considered in the evaluation measurement.\footnote{
Due to lack of space, here we focus on the formal properties of the metrics and we provide references to the definition and explanation 
of the metrics.} 
The upper part of Table~\ref{tab:constraints} summarizes the properties for the retrieval effectiveness metrics.

\input{04_analysis_metrics_tbl}

The  constraints defined by {\citet{Amigo_2013}} assume that relevance judgments are binary. 
However, our axiomatic framework defines the constraints 
\Pri and \Deep over graded relevance (Eq.~\ref{eq:Pri} and~\ref{eq:Deep}, respectively).
Therefore, RR, AP and 
P@$k$ become undefined.\footnote{{\citet{Amigo_2013}}'s analysis shows 
that P@$k$ does not satisfy the \Pri and \Deep constraints, given that it does not consider the order of documents before position $k$.}

The rest of the analysis is inline with the one presented by {\citet{Amigo_2013}}: 
The other metrics (nDCG@$k$,ERR@$k$ and RBP) satisfy \Pri and \Deep constraints 
by applying a relevance discounting factor depending on the depth of the ranking position.
With regards to \DeepTh (Eq.~\ref{eq:DeepTh}) and \CloseTh (Eq.~\ref{eq:CloseTh}) constraints, 
metrics that rewards relevance in deep ranking positions such as AP or nDCG@$k$ satisfy \CloseTh but not \DeepTh, while metrics
that focus on the top of the ranking (P@k, RR and ERR@k) satisfy
\DeepTh but not \CloseTh. RBP satisfies both \CloseTh and \DeepTh.
The reason is that RBP is supported by a  probabilistic  user behavior model that takes into account
the limitations of the ranking exploration process (i.e., user's patience). 
None of these metrics satisfy \Conf.

This family of metrics are not applicable in the context of multiple
query aspects. Therefore, they do not satisfy the diversity-oriented constraints.

\subsection{Intent-Aware Metrics}
\label{sec:analaysis:intent-aware}

The \emph{intent-aware} scheme~\cite{Agrawal-09} extends standard metrics such as AP or ERR to make them applicable to diversification scenarios.
Firstly, each query aspect is evaluated independently and then a weighted average considering query aspect weights is computed. Being  $M_t(\vec{d})$  the score of $\vec{d}$ 
according to the metric $M$ when only the relevance to aspect $t$ is considered:
\begin{equation*}
M\mbox{-IA}(\vec{d}) = \sum_{ t \in \mathcal{T}} w(t) M_t(\vec{d})
\end{equation*}

The central part of Table~\ref{tab:constraints} includes the properties for the intent-aware version of the metrics discussed before. Intent-aware metrics converge to the corresponding standard effectiveness metric when the query has only one aspect. Consequently, they inherit the properties of the  original metric over the relevance-oriented constraints \Pri, \Deep, \DeepTh and \CloseTh. 

Let us now analyze the diversification-oriented constraints. Besides AP-IA@$k$,  RR-IA and P-IA@$k$, which are undefined in the context of graded  relevance judgments, the intent-aware metrics nDCG-IA@$k$, ERR-IA@$k$ and RBP-IA satisfy the \Nov constraint.  If a  document is relevant for several aspects, then the averaged score across query aspects increases.

Most of metrics do not satisfy  \Red and \MRed.
In the case of P-IA@$k$, the precision averaged across aspects in a certain cutoff $k$ is independent from to which particular aspect the documents are relevant to.\footnote{For instance, being $n_i$ the amount of
relevant documents for the aspect $t_i$, the average P@$k$ across 
aspects is: $\frac{1}{|{\mathcal T}|}\sum_{t_i\in{\mathcal T}}\frac{n_i}{k}\propto \sum_{t_i\in{\mathcal T}} n_i$.}
RR-IA@$k$ neither satisfies \Red given that is sensitive only to the first relevant document for each query aspect.  In the case of AP-IA@$k$, 
the relevance contribution of a document to the aspect $t$ is 
higher if relevant documents for $t$ have been observed earlier in the 
ranking.\footnote{The contribution
of a relevant document in AP is proportional
to the precision achieved at the document's position, which
is higher when relevant documents appear in the
previous positions. For instance, being $N_r$ the fixed
amount of relevant documents for every aspect in the collection, and being
$d_{t}$, $d'_{t}$ two documents related with aspect $t$, and $d_{t'}$ a document
related with aspect $t'$ then: $\mbox{AP-IA@}2(d_t,d'_t)=1\frac{1}{N_r}+1\frac{2}{N_r}>
1\frac{1}{N_r}+\frac{1}{2}\frac{1}{N_r}=\mbox{AP-IA@}2(d_t,d_{t'})$}
nDCG-IA@$k$ and RBP-IA also fail to satisfy the \Red constraint. These two metrics are not sensitive to the relevance of 
previously observed documents. The contribution of documents depends on
the rank position and the amount of relevant documents
in the collection. 

 On the other hand,  
the metric ERR-IA@$k$ satisfies both \Red and \MRed,
due to the component $\prod_{j<i}(1-r(d_j,t))$ which estimates the probability of the user
to be satisfied by previously observed documents according to graded relevance levels. 

The \Sat constraint is not satisfied by P-IA@$k$, AP-IA@$k$, nDCG-IA@$k$ nor RBP-IA. The reason is that
all these metrics reward new relevant documents regardless the the gain obtained by previous observed documents. However, the saturation relevance  for RR-IA@$k$ and ERR-IA@$k$ 
 is $1$. Finally, the \AspRel constraint by all the intent-aware metrics analyzed in this work, 
 given that they all consider the first relevant 
document for each aspect in the ranking and all of them consider aspect weights $w(t)$.

\subsection{Other Diversity Metrics}

Besides the intent-aware metrics ($M\mbox{-IA}$), other metrics have been proposed to evaluate the effectiveness of search result diversification systems~\cite{santos2015search}.
{\citet{Zhai-03}} proposed Subtopic Recall (S-Recall@$k$), which measures the number
of aspects captured in the first $k$ positions. Given that the metric only measures the coverage of aspects,  does not satisfy  \Pri, \Deep, \CloseTh and \Conf relevance-oriented constraints. The only diversity oriented constraint that satisfies is \Sat, given that  S-Recall@$k$ considers only the first relevant document for each query aspect and it does not consider aspect weights. Likewise, the metric S-RR@$100\%$ -- an extension to RR  also proposed by~{\citet{Zhai-03}}, defined as the inverse of the rank position 
on which a complete coverage of aspects is obtained -- satisfies the same properties as S-Recall@$k$.

{\citet{Clarke-08}} proposed  Novelty-Biased 
Discounted Cumulative Gain ($\alpha$-nDCG@$k$).\footnote{Note that given that the proposed formal constraints and experiments in this work compare metrics at topic (or query) level, the normalization factor in metrics such as $\alpha$-nDCG@$k$ can be ignored.} This metric is defined as:

\begin{equation*}\label{eq:alphandcg}
\mbox{$\alpha$-nDCG}@k (\vec{d}) =
\sum_{i=1}^{k} 
\frac{\sum_{t\in\mathcal{T}}r(d_i,t)(1-\alpha)^{c(i,t)}}
{\log(i+1)}
\end{equation*}
where $c(i,t)$ represents the amount of documents previously
observed that capture the aspect $t$.
Similarly to the original 
nDCG, it  satisfies  \Pri, \Deep and \CloseTh constraints. However, unlike nDCG,
\DeepTh is also satisfied due to the redundancy factor $(1-\alpha)^{c(i,t)}$, which also
allows to satisfy \Red. \Nov is satisfied due to the additive relevance across aspects. 
In contrast,  $\alpha$-nDCG@$k$ does not satisfy the constraints \MRed and \Sat.  The reason is that the redundancy component $(1-\alpha)^{c(i,t)}$ does not consider the relevance grade of previously observed documents. 
Finally, this metric does not consider the weight of aspects and therefore \AspRel is not satisfied.

{\citet{Clarke-09}} proposed 
Novelty- and Rank-Biased Precision (NRBP), and adaptation of RBP for search 
result diversification, defined as:
\begin{equation*}
\mbox{NRBP}(\vec{d}) = \sum_{i=1}^{\infty}p^{i-1}\sum_{t\in\mathcal{T}} r(d_i,t)(1-\alpha)^{c(i,t)}
\end{equation*}
Similarly to the original RBP, NRBP satisfies
all relevance-oriented constraints except \Conf, given that only relevant documents affect 
the score. In terms of diversity-oriented constraints, NRBP behaves similarly to $\alpha$-nDCG@$k$ given that diversification is modeled in a similar manner.
{\citet{sakai2011evaluating} proposed the D\#-Measure which combines a D-Measure (e.g., D-nDCG~\cite{Sakai-10}) with the ratio of aspects captured in the first 
$k$ positions (modeled by S-Recall@$k$):
\begin{equation*}
\mbox{D\#-Measure}@k(\vec{d})=\lambda \cdot \mbox{S-Recall}@k(\vec{d})+(1-\lambda)\cdot\mbox{D-Measure}@k(\vec{d})
\end{equation*}

NRBP inherits the properties from nDCG-IA@$k$, which already satisfies \DeepTh and \AspRel. Therefore, the S-Recall@$k$ component does not contribute with any additional
 constraint satisfaction.

None of previous metrics satisfy \Conf. However, there exist in the literature \emph{utility}-oriented metrics that penalyze non-relevant documents at the end of the ranking. Two examples are the Normalized Discounted Cumulated Utility (nDCU)~\cite{Yang-07}, and the generalized version Expected Utility (EU)~\cite{Yang-09}. EU is very similar to $\mbox{$\alpha$-nDCG}@k (\vec{d})$ but includes a cost factor. Being $e$ the estimated effort for accessing one document, EU can be expressed as:
\begin{equation*}\label{eq:eu}
\mbox{EU} (\vec{d}) =
\sum_{i=1}^{|\vec{d}|}\frac{1}{1+\log(i)}\left( 
\sum_{t\in\mathcal{T}}r(t) r(d_i,t)(1-\alpha)^{c(i,t)}-e\right)
\end{equation*}
EU inherits the $\alpha\mbox{-nDCG}@k (\vec{d})$ properties, but capturing  \AspRel and \Conf. However EU does still not satisfy \MRed and \Sat.

The Cube Test metric (CT@$k$)~\cite{Luo-13} satisfies  \Sat by adding a saturation factor. Assuming a linear time effort w.r.t. the amount of inspected documents, CT@$k$ can be expressed as:
\begin{equation*}
\mbox{CT@}k( \vec{d})
  = \sum_{i=1}^{|\vec{d}|} \frac{1}{i} 
    \sum_{t\in\mathcal{T}} r(t) r(d_i,t) (1-\alpha) ^ {c(i,t)} f_{\mbox{\it Sat}}
\end{equation*}
where $f_{\mbox{Sat}}$ is 0 or 1 depending if the sum of relevance of documents for the aspect exceeds a certain saturation level. The reciprocal rank discounting factor $\left(\frac{1}{i}\right)$ affects the constraint \CloseTh, rewarding the positions of documents over the amount of relevant documents in top area. In addition, \Conf is neither satisfied. There is no contribution or penalty for documents with zero relevance.

Table~\ref{tab:constraints} also includes the proposed metric \textit{Rank-Biased Utility} ($\mbox{RBU}$), which we describe below.

%% file: 04_analysis_metrics_tbl.tex
\begin{table*}[htp]
\centering
\mycaption{Properties (\tick = constraint satisfied, \fail = constraint not satisfied) of existing retrieval and diversity effectiveness metrics.\label{tab:constraints}}
\begin{adjustbox}{max width=0.6\textwidth}
\begin{tabular}{l ccccc c ccccc}
\toprule
   \multirow{2}{*}{\it Metric}     & \multicolumn{5}{c}{\it Relevance-Oriented Constraints} &  & \multicolumn{5}{c}{\it Diversity-Oriented Constraints} \\
        \cmidrule{2-6}\cmidrule{8-12}
  & \emph{\Pri}  & \emph{\Deep} & \emph{\DeepTh} & \emph{\CloseTh} & \emph{\Conf} &  & \emph{\Nov} & \emph{\Red} & \emph{\MRed} & \emph{\Sat} &   \emph{\AspRel} \\
\midrule
P@$k$ & \fail & \fail & \tick  &  \tick & \fail & & \fail &  \fail &  \fail &  \fail &  \fail  \\
RR & \fail & \fail & \tick  &  \fail & \fail & & \fail &  \fail &  \fail &  \fail &  \fail  \\ 
AP & \fail & \fail & \fail  &  \tick & \fail & & \fail &  \fail  &  \fail &  \fail &  \fail  \\ 
nDCG@$k$ & \tick & \tick & \fail  &  \tick & \fail & &\fail &  \fail &  \fail &  \fail &  \fail  \\
ERR@$k$ & \tick & \tick & \tick  &  \fail & \fail & & \fail &  \fail  &  \fail &  \fail &  \fail  \\  
RBP & \tick & \tick & \tick  &  \tick & \fail & &\fail &  \fail &  \fail &  \fail &  \fail  \\ 
\midrule
P-IA@$k$ & \fail & \fail & \tick  &  \tick & \fail & &\fail &  \fail &  \fail &  \fail &  \tick  \\
RR-IA@$k$ & \fail & \fail & \tick  &  \fail & \fail & & \fail &  \fail &  \fail &  \tick &  \tick  \\  
AP-IA & \tick & \tick & \fail  &  \tick & \fail & & \fail &  \fail  &  \fail &  \fail &  \tick  \\ 
nDCG-IA@$k$ & \tick & \tick & \fail  &  \tick & \fail & &\tick &  \fail &  \fail &  \fail &  \tick  \\ 
ERR-IA@$k$ & \tick & \tick & \tick  &  \fail & \fail & & \tick &  \tick  &  \tick &  \tick &  \tick  \\
RBP-IA & \tick & \tick & \tick  &  \tick & \fail & & \tick &  \fail &  \fail &  \fail &  \tick  \\ 
\midrule
S-Recall@$k$ & \fail & \fail & \tick  &  \fail & \fail & & \fail  &  \fail &  \fail &  \tick &  \fail  \\
S-RR@$100\%$ & \fail & \fail & \tick  &  \fail & \fail & &\fail &  \fail &  \fail &  \tick &  \fail  \\ 

NRBP & \tick & \tick & \tick  &  \tick & \fail & &\tick &  \tick &  \fail &  \fail &  \fail  \\ 
D\#-Measure@$k$& \tick & \tick & \fail  &  \tick & \fail & & \tick &  \fail &  \fail &  \fail &  \tick  \\
$\alpha$-nDCG@$k$ & \tick & \tick & \tick  &  \tick & \fail & & \tick &  \tick &  \fail &  \fail &  \fail  \\ 
EU & \tick & \tick & \tick  &  \tick & \tick & & \tick &  \tick &  \fail &  \fail &  \tick  \\  
CT@$k$ & \tick & \tick & \tick  &  \fail & \fail & & \tick &  \tick &  \fail &  \tick &  \tick  \\  
\midrule
$\mbox{RBU}$@$k$ & \tick & \tick & \tick  &  \tick & \tick & & \tick &  \tick &  \tick &  \tick &  \tick  \\ 
\bottomrule
\end{tabular}
\end{adjustbox}
\end{table*}

%% file: 05_proposal.tex
\section{Rank-Biased Utility}
\label{sec:rbu}

The quality of a diversified ranking depends (at least) on the following factors: (i)~the position of relevant documents in the ranking; (ii)~the redundancy regarding each of the aspects covered by previously observed documents;
(iii)~the weights of the aspects seen in the ranking and (iv) the \emph{effort} -- in terms of user cost or time -- derived from inspecting relevant or non-relevant documents. The analysis described in Section~\ref{sec:metric:analysis} shows that none of the existing metrics take into account all these factors.
To fill this gap, we propose \textit{Ranking-Biased Utility} (RBU), which satisfies all the retrieval and diversity-oriented formal constraints (see proofs in the appendix).

The analysis shows that RBP~\cite{RBP} is the only metric that satisfies the four 
first relevance constraints, while ERR-IA@$k$~\cite{Agrawal-09, chapelle2009expected} is the only metric that satisfies all the five diversity-oriented constraints. Expected Utility (EU) is the only that satisfies \Conf,  capturing the suitability of the ranking cutoff. 

In order to satisfy every constraint, RBU combines the user 
exploration deepness model from RBP with the redundancy modeled in ERR-IA@$k$, and also adds the \emph{user effort} component $e$ in EU to satisfy the \Conf constraint.

 The metrics RBP and ERR-IA@$k$ can be combined together under the following user behavior assumptions: 
(i)~The user has a probability
$p$ to explore the next document and (ii)~the user has a probability
$r(d_j,t)$ to get gain from document $d_j$ for the topic $t$. 

Similarly to the ERR-IA@$k$, the probability of being satisfied by document 
$d_i$ after observing the documents that occur earlier in the ranking is:
$$
r(d_i,t)\prod_{j=1}^{i-1}(1-r(d_j,t))
$$
Analogously to the user model followed by RBP, the resulting 
contribution of a document $d_i$ in the position $i$ must be weighted 
according to $p^i$:
\begin{equation*}
p^i r(d_i,t) \prod_{j=1}^{i-1}(1-r(d_j,t))
\end{equation*}

In order to satisfy \AspRel, the weighted sum of contributions across aspects in $\mathcal{T}$ is:
\begin{equation*}
p^i \sum_{t\in\mathcal{T}} w(t) r(d_i,t) \prod_{j=1}^{i-1}(1-r(d_j,t))
\end{equation*}

And the cumulative gain across rank positions until $k$ is:
\begin{equation*}
\mbox{RBU}@\mbox{k}(\vec{d})=\sum_{i=1}^k p^i \sum_{t\in\mathcal{T}}w(t)r(d_i,t)\prod_{j=1}^{i-1}\left(1-r(d_j,t)\right)
 \end{equation*}

Similarly to EU, we  define RBP in utility terms in order to capture \Conf. 
Being $e$ the effort of observing a document,
the rank biased accumulated effort is weighted according to $p^i$, that is: $ \left(\sum_{i=1}^k p^i e \right)$.

Finally, combining the relevance contribution with the cumulative effort, we obtain:
\begin{equation}\label{eq:rbu}
\resizebox{0.9\hsize}{!}{$\displaystyle
 \mbox{RBU$@$k}(\vec{d})=\sum_{i=1}^kp^i \left( \sum_{t\in\mathcal{T}}\left(w(t) r(d_i,t)\prod_{j=1}^{i-1}\left(1-r(d_j,t)\right) \right) - e \right)
 $}
 \end{equation}

RBU@$k$ matches with the  RBP-IA metric when assuming a zero
effort ($e=0$), and a small contribution of documents in terms of gain for query aspects, 
$$r(d_i,t)\ll1\Longrightarrow\prod_{j=1}^{i-1}\left(1-r(d_j,t)\right)\simeq 1\Longrightarrow $$

$$ \mbox{RBU$@$k}(\vec{d})= \sum_{t\in \mathcal T} w(t) \sum_{j\le i}  \left(p^{i-1} r(d_i,t) 1 \right) - 0 = \sum_{t\in\mathcal{T}} w(t) \hspace{0.1em}\mbox{RBP}_t(\vec{d})$$

On the other hand, RBU@$k$ is equivalent to the metric ERR-IA@$k$ when the effort component is zero ($e=0$), and the probability of exploring the next document is maximal ($p=1$):
$$\sum_{i=1}^k 1^i \left( \sum_{t\in\mathcal{T}} \left( w(t) r(d_i,t) \prod_{j=1}^{i-1}(1-r(d_j,t) ) \right) -0 \right)  = \sum_{t\in\mathcal{T}} w(t) \hspace{0.1em}\mbox{ERR}_t@k(\vec{d})$$

We now discuss the role of the effort component $e$, which 
represents the cost inherently associated to inspect a new document in the ranking.
\footnote{In this work, the effort of inspecting or judging a relevant or non-relevant document is the same. We leave for future work the definition of formal constraints that consider these differences~\cite{turpin2009including, Smucker-12}.}
For instance, if $e=0.1$ and the inspected document $d_i$ has a relevance of $0.1$ to aspect $t_i$, then the actual gain is zero:
$$r(d_i,t)\prod_{j<i}\left(1-r(d_j,t)\right)-e=
0.1\prod_{j<i}\left(1-0\right)-0.1=0$$

We have introduced RBU@$k$ and shown that the proposed metric satisfies all the relevance- and diversity-oriented formal constraints. The experiments described in the following sections  compare RBU@$k$ to other metrics in the context of standard evaluation campaigns for search result diversification. 

%% file: 06_experiments.tex
\section{Experiments}
\label{sec:metaevaluation}

We start defining our meta-evaluation metric. Then we evaluate the metrics 
in different scenarios based on the TREC Web Track 2014 ad-hoc retrieval 
task~\cite{collins2015trec}, which includes search result diversification. Finally, we corroborate our results under the context of  the TREC 
Dynamic Domain task~\cite{yang2015overview}.\footnote{Releasable data and scripts used in these experiments are available at \url{https://github.com/jCarrilloDeAlbornoz/RBU}. Diversity metrics and RBU are also included in the EvALL evaluation framework~\cite{amigo2017evall} \url{http://evall.uned.es/}.}

\subsection{Meta-evaluation: Metric Unanimity}

We aim to quantify the ability of metrics to
capture diversity in addition to traditional ranking quality aspects.
For this purpose,  we define the \emph{Metric Unanimity (\MU)}. 
\MU quantifies to what extent a metric is sensitive to quality
aspects captured by other existing metrics. It follows a similar concept used by \textit{Strictness},\footnote{Strictness checks to what extent  a metric can outscore metrics that achieve a low score according to other metrics.} proposed by~{\citet{Amigo_2013}} for the ad-hoc retrieval scenario.

Our intuition is that, if a system improves another system for every quality criteria, this should be \emph{unanimously} reflected by every metric. A metric that captures all quality criteria should
 reflect these improvements.
\input{06_experiments_scenarios_tbl}

Considering the space of system output pair comparisons (i.e., $Q\vec{d})>Q(\vec{d}')$) and a set of metrics, \MU can be formalized as  the Point-wise Mutual Information (PMI) between decisions of a metric and 
 improvements reported simultaneously by the rest of metrics in the set. Formally, let be $m$ a metric, $\mathcal{M}$ the rest of metrics, and a set of system outputs $\mathcal{S}$. Being  $\Delta m_{i,j}$ and $\Delta{\mathcal M}_{i,j}$ statistical variables over  system pairs $(\vec{d_i},\vec{d_j}) \in \mathcal{S}^2$, indicating a system improvement  according to the metric and to the rest of metrics, respectively: \footnote{The a priori probability of a system improvement for every metric is fixed $P(\Delta m_{i,j})=\frac{1}{2}$. That is, for the cases on which two system outputs obtain the same score $m(\vec{d_i})=m(\vec{d_j})$, we add $0.5$ to the statistical count.}
\begin{align*}
&\Delta m_{i,j}\equiv m(\vec{d_i})> m(\vec{d_j})\\
&\Delta{\mathcal M}_{i,j}\equiv\forall m\in \mathcal{M} \ldotp\left(m(\vec{d_i})\ge m(\vec{d_j})\right)
\end{align*}

Then \MU is formalized as:

\begin{equation*}\label{eq:um}
\resizebox{.9\hsize}{!}{%
$\mbox{MU}_{\mathcal{M},\mathcal{S}}(m) = \mbox{PMI} \left( 
\Delta m_{i,j}, \Delta {\mathcal M_{i,j}} \right)
=
\log \left( 
   \frac
   {P(\Delta m_{i,j},\Delta {\mathcal M_{i,j}})}
   {P(\Delta m_{i,j})\cdot P(\Delta {\mathcal M_{i,j}})}     
    \right)$
    }
\end{equation*}

Let us consider the following example illustrated by the Table below:
\begin{center}
\begin{tabular}{ c c c c }
\toprule
& $m^1$ & $m^2$ & $m^3$\\
\midrule
$S_1$  & 1 & 0.8 & 1 \\
 $S_2$ & 0.5 & 0.3 & 0.2 \\
 $S_3$ & 0.2 & 0.4 & 0.5 \\
\bottomrule
\end{tabular}
\end{center}
\vspace{0.6em}
 The example consists of three metrics and three system outputs. We now compute the \MU of the metric $m^1$ regarding the rest of metrics $\mathcal{M}=\{m^2, m^3\}$.  Here, there are 6 sorted pairs of system outputs: $(S_1,S_2)$,$(S_2,S_1)$, $(S_1,S_3)$, etc. The improvements reported by $m^1$ are: $\Delta m^1_{1,2}$, $\Delta m^1_{1,3}$, and $\Delta m^1_{2,3}$.  The improvement reported simultaneously by the other metrics  are: $\Delta \mathcal{M}_{1,2}$, $\Delta \mathcal{M}_{1,3}$, and $\Delta \mathcal{M}_{3,2}$.  $m^1$ agrees with $\mathcal{M}$ in two cases. Therefore $\MU_{\mathcal{M}}(m^1)=\log\left(\frac{2/6}{3/6\cdot 3/6}\right)=0.415$.

\MU has four properties that we describe below.
\begin{description}
\item[{\it Property 1.}] Capturing  every unanimous improvement maximizes \MU  regardless the other
 decisions:
$$\mbox{MU}_{\mathcal{M},\mathcal{S}}(m)=\log\left(\frac{P(\Delta m_{i,j},\Delta {\mathcal M_{i,j}})}{\frac{1}{2}\cdot k}\right)\propto P(\Delta m_{i,j},\Delta {\mathcal M_{i,j}})$$
\item[{\it Property 2.}] A metric $m_{\mbox{\it rand}}$  which assigns random or constant scores 
to every system outputs achieves a zero \MU, capturing the \emph{sensitivity} of metrics:
$$
\mbox{MU}_{\mathcal{M},\mathcal{S}}(m_{\mbox{\it rand}})
= \log \left( 
   \frac
   {\frac{1}{2}\cdot P(\Delta {\mathcal M_{i,j}})}
   {\frac{1}{2}\cdot  P(\Delta {\mathcal M_{i,j}})}
   \right)
= \log(1) = 0
$$
\item[{\it Property 3.}] \MU is asymmetric. A metric $m$ can be \textit{unanimous} regarding the rest of metrics,
while the rest of metrics are not.  
$$MU_{\{m_2,m_3\}}(m_1) \neq MU_{\{m_1,m_3\}}(m_2)\neq  MU_{\{m_1,m_2\}}(m_3)$$
\item[{\it Property 4.}] \MU is not affected
by the predominance of a certain family of metrics in the set $\mathcal{M}$:
$$
\mbox{MU}_{\mathcal{M}\cup\{m'\},\mathcal{S}}(m)
=
\mbox{MU}_{\mathcal{M}\cup\{m',m',\ldots, m'\},\mathcal{S}}(m)
$$
\end{description}

\subsection{Experiment 1: TREC Web Track 2014}
\label{sec:experiments:official}
This first experiments aims to measure \MU in a standard diversification evaluation campaign: the TREC Web Track 2014 ad-hoc retrieval task~\cite{collins2015trec}. In this benchmark, systems need to perform ad-hoc retrieval from the ClueWeb-12 collection, for a total of 50 test topics and return the top 10,000 documents. Some of the topics have multiple aspects --therefore, diversified rankings may be more effective. We use the 30 official runs submitted to the ad-hoc retrieval task and available at TREC's website. 

Using our own implementation of the metrics,  we execute over the official runs the following metrics:
AP, RR, AP-IA and RR-IA which do not require any parameter;
 P@$k$,  ERR@$k$, NDCG@$k$ and their corresponding intent-aware variants,
using $k \in \{10, 20, 50,$  $100, 1000\}$; 
S-Recall@$k$, RBP, NRBP and $\alpha$-nDCG@$k$;  with $p\in\{0.8,0.9,0.99\}$
} and $\alpha\in\{0.1,0.25,0.5,0.75\}$; 
EU and our proposed metric RBU with the effort parameter  $e \in \{0.001,0.05,0.1,0.5\}$.

 For metrics that do not accept multiple query aspects, we consider the maximum relevance  across aspects: $r(d)=max_{t\in{\mathcal T}}(r(d,t))$.

The first  column in  Table~\ref{tab:results} shows the metrics ranked by \MU. For the sake of clarity, the table includes for each metric the variant with highest \MU.
Results show that metrics that satisfy only a few constraints
such as P@$k$ or S-Recall@$k$ are substantially less unanimous than
the rest of metrics. This means that metrics with higher scores cover the same quality criteria captured by P@$k$ or S-Recall@$k$, but these two metrics do not capture other criteria captured
by the rest of metrics.

Our second observation is that a metric with a shallow cutoff (e.g., ERR@$50$) -- i.e., it takes into account a few documents in the ranking -- has lower \MU score than its deep counterpart (e.g., ERR@$1000$). This behavior is consistent for every metric and variants. Likewise, higher values for the patience parameter $p$ in RBP obtains higher \MU scores. Intuitively, the shallower the metric is, the less probable is to capture improvements in deep ranking positions.

RBU obtains the highest scores, when $p=0.99$ (i.e., the metric considers deep positions in the ranking) and all the tested values for the effort component $e$.

\input{06_experiments_official_tbl}

\subsection{Experiment 2: Simulating  Alternative Scenarios}
\label{sec:experiments:simulations}

In order to study the behavior of metrics under different situations and to corroborate our findings, we repeat the experiment described before after artificially modifying some parameters of the official TREC Web Track experimental setup.

The second column in Table~\ref{tab:results}
shows the results when: 
\begin{enumerate}
\item Enforcing all relevance judgments to be graded: we  replace each 
discrete relevance value $r$ by a random value between zero and $r$ 
: $r'(d)=\mbox{\it rand}(0..r(d))$. This is related to the \MRed constraint.

\item Randomly assigning a certain weight to each aspect $t$ in such a way that the sum 
of the weights for each topic (or query) adds up to 1: $w(t)=\mbox{\it rand}(0..1)$ and $\sum_{t \in {\mathcal T}}w'(t)=1$. This is related to the \AspRel constraint.

\item The ranking of documents returned by each system is manipulated by reducing randomly its length: $|\vec{d}|=\mbox{rand}(0,\ldots,|\vec{d}|)$.
This variation simulates the situation in which
  systems should cut their output rankings
according to their confidence of retrieving (or not) more relevant documents.
This tuning is related to the \Conf constraint, which is only satisfied by EU and the proposed metric. 
\end{enumerate}

As a result, the difference in terms of \MU scores between RBU and the other metrics is larger in this simulated scenario. The experiment suggests that this effect is not due to the fact of satisfying any single constraint, but satisfying several constraints simultaneously.
Although EU satisfies \Conf and ERR-IA@$k$ satisfies \MRed and \Sat, RBU outperforms both metrics in terms of \MU.

In all the previous experiments, we have seen that \MU rewards
the fact of considering deeper positions in the ranking. In order
to isolate this variable, the next simulation (Table~\ref{tab:results}, third column) reduces the length of rankings substantially, 
by defining a random cutoff between $0$ and $50$: $|\vec{d}|=\mbox{rand}(0..50)$. Consequently,
 metrics that use a cutoff equal or greater than $k=50$ will not be rewarded by \MU. Remarkably, all the RBU variants with an effort parameter $e$ higher than zero obtain the highest \MU scores -- RBU with $e=0$ (omitted in the table) achieves a 0.7709 \MU score. 

This suggests that the effort component $e$ plays an important role when evaluating rankings with different lengths.

\subsection{Experiment 3: Considering Metrics and  Default Parameters used in Official Evaluation}

\MU scores depend on the set of metrics in consideration. Therefore, the results could be biased by the selected metric set $\mathcal{M}$ and variants. In order to avoid this bias, we consider the official metrics and parameters used by the 
TREC Web Track organizers. In addition, to avoid the effect of implementation variations or bugs, we compare RBU (implemented by ourselves) against the official evaluation scores 
released by TREC (first column in Table~\ref{tab:results2}).

In this case, AP-IA gets the highest \MU score. In terms of RBU, we can see that $p$ values and \MU scores are correlated. This shows again than \MU is biased by the the amount of documents in the ranking that are \emph{visible} to the metric. Note that most of metrics proposed by the organizers use a cutoff no greater than $k=20$. That is, most of metrics
receive less information than AP-IA or NRBP, which take into account \emph{all} the documents in the ranking.

In order to avoid this effect, we focus on metrics that apply the the cutoff $k=20$, and we apply the same cutoff to RBU: RBU@$20$\footnote{In this experiment
we use the official evaluation scores. Therefore, we cannot adapt
AP-IA nor NRBP to this cutoff.}
 Maintaining the amount of documents visible to metrics constant, RBU achieves the same \MU score (0.9556) for all the tested variants, obtaining the highest \MU score among the metrics. This suggests that the RBU performance in terms of \MU is not due to differences in the length of the observed ranking.

The high \MU scores of RBU could be possibly due to the fact of having an explicit component for the user effort ($e$ parameter),
rather than the ability to capture other quality aspects such as diversity and redundancy.
In order to isolate this variable, we consider only three RBU variants with zero value
in the effort parameter $\left(e=0, p=\left\{0.8,0.9,0.99\right\}\right)$. Results at the bottom of second column in Table~\ref{tab:results2} show that RBU also outperforms the rest of metrics when $e=0$.

\subsection{Experiment 4: Validation using TREC Dynamic Domain Track}

In order to check the robustness of our
empirical conclusions, we repeat the same experiment over
TREC Dynamic Domain 2015~\cite{yang2015overview}, which includes 23 official runs.
This track consists of an interactive search scenario. 
Systems receive aspect-level feedback iteratively and need to 
dynamically retrieve as many relevant documents for aspects 
as possible, using as few iterations as possible. 
An important particularity of this task is that the system
must predict the optimal ranking cutoff which is
 closely related with the \Conf constraint.
The official metrics used in this track are Cube Test (CT@$k$) and 
Averaged Cube Test (ACT@$k$)~\cite{Luo-13}, which are included in our experiments.

The rightmost column in Table~\ref{tab:results2} shows that we obtain similar results: all the RBU variants are at the top of the metrics ranking. In this case, the 
user effort parameter $e$ is important, given that it is necessary 
to outperform other metrics such as CT@$k$ or ACT@$k$. In addition, we achieved again the same
result when considering only one RBU variant, appearing at
the top in terms of \MU scores.

%% file: 06_experiments_scenarios_tbl.tex
\begin{table*}
\mycaption{Metric Unanimity scores (\MU) for the TREC Web Track 2014 ad-hoc retrieval task: official (Section~\ref{sec:experiments:official}) and simulated scenarios (Section~\ref{sec:experiments:simulations}). Given that normalization has not effect in terms of formal constraints and \MU, which work at topic (query) level, normalized version of metrics behave similarly to the metric without normalization \mbox{(e.g., $\mbox{MU}(\mbox{nDCG}) = \mbox{MU}(\mbox{DCG})$)} and therefore are not included.\label{tab:results}}
\begin{adjustbox}{max width=0.928\textwidth}
\renewcommand{\tabcolsep}{0.25em}
\begin{tabular}{lc l 
                lc l lc}
\toprule

\multicolumn{2}{c}{\multirow{4}{*}{Official}} & & \multicolumn{5}{c}{Simulated Scenarios} \\

\cmidrule{4-8}
&&
&\multicolumn{2}{c}{$r'(d)=\rand(0,r(d))$} &
&\multicolumn{2}{c}{ $r'(d)=\rand(0,r(d))$}\\

&&
&\multicolumn{2}{c}{$r'(t)=\rand(0,r(t))$}&
&\multicolumn{2}{c}{$r'(t)=\rand(0,r(t))$}
\\

&&
&\multicolumn{2}{c}{$|\vec{d}|=\rand(0,|\vec{d}|)$} &
&\multicolumn{2}{c}{$|\vec{d}|=\rand(0,50)$}
\\

\cmidrule{1-2}
\cmidrule{4-5}
\cmidrule{7-8}

{\bf  RBU$_\mathbf{e=\{0.001,0.05,0.1,0.5\},p=0.99}$} & 0.8024 &
&{\bf RBU$_\mathbf{e=\{0.001,0.05,0.1,0.5\},p=0.99}$} & 0.8568 &
& {\bf RBU$_\mathbf{e=\{0.001,0.05,0.1,0.5\},p=\{0.8,0.9,0.99\}}$} & 0.9808\\

$\alpha$-DCG-IA@$1000_{\alpha=\{0.1,0.25,0.5\}}$ & 0.7956 &
& $\alpha$-DCG-IA@$1000_{\alpha=\{0.1,0.25,0.5,0.75\}}$ & 0.7734 &
& $\alpha$-DCG-IA@\{$50$,$100$,$1000$\}$_{\alpha=\{0.1,0.25,0.5,0.75\}}$ & 0.7709\\

DCG@$1000$ & 0.7956 &
& DCG@1000 & 0.7734 &
& DCG-IA@\{50,100,1000\} & 0.7709\\

DCG-IA@1000 & 0.7956 &
& DCG-IA@1000 & 0.7734 &
& EU$_{\alpha=\{0.1,0.25,0.5,0.75\},e=\{0,0.001,0.05,0.5\}}$ & 0.7709\\

EU$_{\alpha=\{0.1,0.25,0.5\},e=\{0,0.05,0.1,0.5\}}$ & 0.7956 &
& EU$_{\alpha=\{0.1,0.25,0.5,0.75\},e=\{0,0.001,0.05,0.5\}}$ & 0.7734 &
& ERR-IA@\{50,100,1000\} & 0.7709\\

ERR-IA@1000 & 0.7956 &
& ERR-IA@1000 & 0.7734 &
& NRBP$_{p=\{0.8,0.9,0.99\},\alpha=\{0.1,0.25,0.5,0.75\}}$ & 0.7709\\

ERR@1000 & 0.7956 &
& ERR@1000 & 0.7734 &
& DCG@\{50,100,1000\}   & 0.7687\\

NRBP$_{p=\{0.8,0.9,0.99\},\alpha=\{0.1,0.25,0.5\}}$ & 0.7956 &
& AP & 0.7734 &
& ERR@\{50,100,1000\} & 0.7679\\

AP & 0.7926 &
& AP-IA & 0.7734 
&& AP-IA & 0.7642\\

AP-IA & 0.7926 &
& NRBP$_{p=\{0.8,0.9,0.99\},\alpha=\{0.1,0.25,0.5,0.75\}}$ & 0.7734 &
& AP & 0.7627\\

RBP$_{p=\{0.8,0.9,0.99\}}$ & 0.7911 &
& RBP$_{p=0.99}$ & 0.7717 &
& RBP$_{p=\{0.8,0.9,0.99\}}$ & 0.7597\\

P-IA@20 & 0.7272 &
& P@\{20,50\} & 0.7103 &
& P-IA@20 & 0.7077\\

P@20 & 0.7192 &
& P-IA@\{20,50\} & 0.7103 &
& P-IA@10 & 0.6888\\

RR-IA & 0.6835 &
& RR-IA & 0.6704 &
& RR-IA & 0.6841\\

RR & 0.6486 &
& RR & 0.6082 &
& RR & 0.6561\\

S-Recall@10 & 0.3965 &
& S-Recall@10 & 0.4238 &
& S-Recall@10 & 0.5137\\

S-Recall@20 & 0.3538 &
& S-Recall@20 & 0.4084 &
& S-Recall@20 & 0.4994\\

S-Recall@50 & 0.3065 &
& S-Recall@50 & 0.3658 &
& S-Recall@100 & 0.4831\\

S-Recall@100 & 0.2478 &
& S-Recall@100 & 0.3007 &
& S-Recall@50 & 0.4831\\

\bottomrule
\end{tabular}
\end{adjustbox}
\end{table*}

%% file: 06_experiments_official_tbl.tex
\begin{table*}
\mycaption{\MU scores over official metrics in TREC Web Track 2014 and TREC Dynamic Domain Track 2015.
}
\label{tab:results2}
\begin{adjustbox}{max width=0.915\textwidth}
\renewcommand{\tabcolsep}{0.25em}
\begin{tabular}{lc c lc c lc}
\toprule
\multicolumn{5}{c}{ TREC Web Track 2014 (Official Metrics)} &
&
\multicolumn{2}{c}{TREC Dynamic Domain 2015 (Official Metrics)}\\

\cmidrule{1-5} \cmidrule{7-8}
\multicolumn{2}{c}{Official} &
&
\multicolumn{2}{c}{$k=20$} 
&
&\multicolumn{2}{c}{Official}\\

\cmidrule{1-2}\cmidrule{4-5}\cmidrule{7-8}

AP-IA & 0.9771 &
& {\bf RBU}$_\mathbf{e=*,p=*}$ & 0.9556  &
& {\bf RBU}$_\mathbf{e=\{0.001,0.05,0.1,0.5\},p=0.99}$ & 0.8488\\

{\bf RBU}$_\mathbf{e=\{0,0.001,0.05,0.1,0.5\},p=0.99}$ & 0.9770..0.9766 & &
\{ $\alpha$-nDCG, $\alpha$-nDCG \}@20 & 0.9427 &&
 {\bf RBU}$_\mathbf{e=0.001,p=0.9}$ & 0.8453\\
 
{\bf RBU}$_\mathbf{e=\{0,0.001,0.05,0.1,0.5\},p=0.9}$ & 0.9763..0.9760 & &
 \{ ERR-IA, nERR-IA \}@20 & 0.9425 &&
  {\bf RBU}$_\mathbf{e=\{0.05,0.1\},p=0.9}$ & 0.8441\\
  
{\bf RBU}$_\mathbf{e=\{0,0.001,0.05,0.1,0.5\},p=0.8}$ & 0.9760..0.9750 &&
 P-IA@20 & 0.9080 & &
 {\bf RBU}$_\mathbf{e=0.5,p=0.9}$ & 0.8440\\
 
\{ $\alpha$-DCG, $\alpha$-nDCG \}@20 & 0.9540 &&
 S-Recall@20 & 0.4141 &&
  {\bf RBU}$_\mathbf{e=0.001,p=0.8}$ & 0.8406\\
  
ERR-IA@20, nERR-IA@20 & 0.9539 &&
  &  & &
  {\bf RBU}$_\mathbf{e=\{0.05, 0.1, 0.5\},p=0.8}$ & 0.8396\\
    
NRBP, nNRBP & 0.9509 & &
 & &  &
   ACT@10 & 0.6276 \\
 
\{ ERR-IA, nERR-IA, $\alpha$-DCG, $\alpha$-nDCG \}@10 & 0.9373 & &
  &  & &
 ERR (Arith. Mean) & 0.5955 \\
  
P-IA@20 & 0.9310 & &
  \multicolumn{2}{c}{$k=20$, $e=0$}    & &
  CT@10 & 0.5938 \\
 
   \cmidrule{4-5}

P-IA@10 & 0.9071 & &
{\bf RBU}$_\mathbf{e=0, p=\{0.8,0.9,0.99\}}$ & 0.9556 & &
  {\bf RBU}$_\mathbf{e=0,p=\{0.8,0.9,0.99\}}$ & 0.5937\\
 
\{ $\alpha$-DCG, $\alpha$-nDCG \}@5 & 0.9001 & &
\{  $\alpha$-DCG, $\alpha$-nDCG \}@20 & 0.9428 & &
 ERR (Harm. Mean) & 0.5912\\
  
\{ ERR-IA, nERR-IA \}@5 & 0.8999 & &
\{  ERR-IA, nERR-IA \}@20 & 0.9425 &&
 P@Recall & 0.1162 \\
  
P-IA@5 & 0.8720 & &
 P-IA@20 & 0.9081 &&
P@Recall (modified) & 0.1044\\
 
S-Recall@5 & 0.5573 & &
 S-Recall@20 & 0.4146 & & 
RR@10 & 0.1031\\
 
S-Recall@10 & 0.5001 &  &
&  & &
& \\

S-Recall@20 & 0.4515 &  &
&  & &
& \\

\bottomrule
\end{tabular}
\end{adjustbox}
\end{table*}

%% file: 07_conclusions.tex
\section{Conclusions}
\label{sec:conclusions}

We defined an axiomatic framework to analyze diversity metrics and found that none of the existing metrics satisfy all the constraints. Inspired by this analysis, we proposed Rank-Biased Utility (RBU, Equation~\ref{eq:rbu}), which satisfies all the formal constraints. Our experiments over standard diversity evaluation campaigns show that the proposed metric has more \emph{unanimity} than the official metrics used in the campaigns, i.e., RBU captures more quality criteria than the ones captured by other metrics. 
We believe our contributions would help researchers and analysts to define their evaluation framework (e.g., which evaluation metric should be used?) in order to analyze the effectiveness of systems in the context of scenarios involving search result diversification. Future work includes a further parameter sensitivity analysis of metrics, as well as the study of other meta-evaluation criteria such as sensitivity or robustness against noise.

%% file: 08_Appendix.tex
\newcommand\eatpunct[1]{}
\renewcommand{\proof}[1]{\paragraph*{\emph{\textsc{Proof.}}} #1}

\renewcommand{\Pri}[0]{\emph{\textsc{Pri}}\xspace}
\renewcommand{\Deep}[0]{\emph{\textsc{Deep}}\xspace}
\renewcommand{\DeepTh}[0]{\emph{\textsc{DeepTh}}\xspace}
\renewcommand{\CloseTh}[0]{\emph{\textsc{CloseTh}}\xspace}
\renewcommand{\Conf}[0]{\emph{\textsc{Conf}}\xspace}

\renewcommand{\Nov}[0]{\emph{\textsc{AspDiv}}\xspace}
\renewcommand{\Red}[0]{\emph{\textsc{Red}}\xspace}
\renewcommand{\MRed}[0]{\emph{\textsc{MRed}}\xspace}
\renewcommand{\Sat}[0]{\emph{\textsc{Sat}}\xspace}
\renewcommand{\AspRel}[0]{\emph{\textsc{AspRel}}\xspace}

\newcommand{\RBU}[0]{$\mbox{RBU}$\xspace}

\section*{Appendix: Formal Proofs}
\begin{proof}
\textit{Rank-Biased Utility (\RBU, Eq.~\ref{eq:rbu}) satisfies the constraints: \Pri (Eq.~\ref{eq:Pri}), \Deep (Eq.~\ref{eq:Deep}), \DeepTh (Eq.~\ref{eq:DeepTh}) and \CloseTh (Eq.~\ref{eq:CloseTh}).}
%
%
{\small
\RBU is defined as:
$$\mbox{RBU$@$k}(\vec{d})=\sum_{i=1}^kp^i \left( \sum_{t\in\mathcal{T}}\left(w(t)r(d_i)\prod_{j=1}^{i-1}(1-r(d_j,t))\right)-e \right)$$
In the context of these constraints, it is assumed that there is only a single aspect $t$ for a given query or topic. Therefore,
\RBU can be expressed as:
$$\mbox{RBU$@$k}(\vec{d})=\sum_{i=1}^kp^i \left( \left(r(d_i)\prod_{j=1}^{i-1}(1-r(d_j,t))\right)-e \right)$$
In addition, the condition \emph{relevance contribution} is assumed, i.e., the relevance
of single documents does not completely cover the user information needs $r(d)\ll 1$. Therefore, we can assume
that $$\prod_{j=1}^{i-1}(1-r(d_j,t))\simeq \prod_{j=1}^{i-1} 1=1$$
Finally, the four constraints compare rankings with the same length. This means
that we can eliminate the user cost component $e$, which is $e\sum_{i=1}^kp^i$ 
for every ranking in comparison. Under all
these assumptions, \RBU is equivalent to the traditional $\mbox{RBP}$ metric~\cite{RBP}:
$$\mbox{RBU$@$k}(\vec{d})\propto\sum_{i=1}^k p^i r(d_i)=\mbox{RBP$@$k}(\vec{d})$$
According to the study by {\citet{Amigo_2013}}, $\mbox{RBP}$ satisfies the four constraints enumerated above.
}
\end{proof}
\begin{proof}
\textit{\RBU satisfies the \Conf constraint (Eq.~\ref{eq:Conf}).}

{\small
\RBU can be expressed as:
{\scriptsize
$$ \mbox{RBU$@$k}(\vec{d})=\sum_{i=1}^kp^i \sum_{t\in\mathcal{T}}\left(w(t)r(d_i)\prod_{j=1}^{i-1}(1-r(d_j,t))\right)-e\sum_{i=1}^kp^i$$
}
then
%
{\scriptsize
\begin{align*}
&\mbox{RBU}@k\left(\vec{d}\right)>\mbox{RBU}@k\left(\vec{d}, d^{\neg rel}\right)\Leftrightarrow\\
&\mbox{RBU$@$k}(\vec{d})>\mbox{RBU$@$k}(\vec{d})-p^{n+1}e\Leftrightarrow 0>-p^{n+1}e \\
\end{align*}
}
}
\end{proof}
\begin{proof}
\textit{\RBU satisfies the \Nov constraint (Eq.~\ref{eq:Nov}).}
%
%
{\small
Under the constraint conditions: 
$\mbox{RBU}\left(\vec{d}_{d_i\leftrightarrow d_i'}\right)>\mbox{RBU}\left(\vec{d}\right)$ is equivalent to:
{\scriptsize
\begin{align*}
&p^i\sum_{t\in\mathcal{T}}\left(w(t)r(d_i',t)\prod_{j=1}^{i-1}(1-r(d_j,t))\right)>
p^i\sum_{t\in\mathcal{T}}\left(w(t)r(d_i,t)\prod_{j=1}^{i-1}(i-r(d_j,t))\right)\Leftrightarrow\\
&\sum_{t\in\mathcal{T}}\left(w(t)r(d_i',t)\right)>
\sum_{t\in\mathcal{T}}\left(w(t)r(d_i,t)\right)\Leftarrow \sum_{t\in\mathcal{T}}\left(r(d_i',t)\right)>
\sum_{t\in\mathcal{T}}\left(r(d_i,t)\right)\Leftarrow\\
&\forall t\in\mathcal{T} \ldotp r(d_i',t)>r(d_i,t)
\end{align*}
}}
\end{proof}
\begin{proof}
\textit{\RBU satisfies the \Red  constraint (Eq.~\ref{eq:Red}).}
%
%
{\small
 Under the constraint conditions:
{\scriptsize
\begin{align*}
\mbox{RBU}
                      \left(\vec{d},d'\right) &> \mbox{RBU}\left(\vec{d},d\right) \Leftrightarrow\\
& w(t')r(d',t')\prod_{j=1}^{|\vec{d}|}(1-r(d_j,t'))>w(t)r(d,t)\prod_{j=1}^{|\vec{d}|}(1-r(d_j,t))\Leftrightarrow\\
&\prod_{j=1}^{|\vec{d}|}(1-r(d_j,t))>\prod_{j=1}^{|\vec{d}|}(1-r(d_j,t))\Leftrightarrow\\
&(1-r_c)^{\left|\left\{d_i\in\vec{d} | r(d_i,t')=r_c\right\}\right|}>(1-r_c)^{\left|\left\{d\in\vec{d} | r(d,t)=r_c\right\}\right|}\Leftrightarrow\\
&\left|\left\{d_i\in\vec{d} | r(d_i,t)=r_c\right\}\right|>\left|\left\{d\in\vec{d} | r(d,t')=r_c\right\}\right|
\end{align*}
}
}
\end{proof}
%
%
\begin{proof}
\textit{\RBU satisfies the \MRed constraint (Eq.~\ref{eq:MRed}).}
%
%
{\small
Under the constraint conditions:
{\scriptsize
\begin{align*}
\mbox{RBU}\left(\vec{d},d'\right)&>\mbox{RBU}\left(\vec{d},d\right)\Leftrightarrow\\
& w(t')r(d',t')\prod_{j=1}^{|\vec{d}|}(1-r(d_j,t'))>w(t)r(d,t)\prod_{j=1}^{|\vec{d}|}(1-r(d_j,t))\Leftrightarrow\\
&\prod_{j=1}^{|\vec{d}|}(1-r(d_j,t))>\prod_{j=1}^{|\vec{d}|}(1-r(d_j,t))\Leftarrow \forall d_i\in \vec{d} \ldotp r(d_i,t)>r(d_i,t')
\end{align*}
}
}
\end{proof}
\begin{proof}
\textit{\RBU satisfies the \Sat constraint (Eq.~\ref{eq:Sat}).}
%
%
{\small
There exists a relevance value $r(d_n,t)=r_{max}=1$
 large enough such that:
{\scriptsize
\begin{align*}
\mbox{RBU}\left(\vec{d},d_{n+1}\right)&=\sum_{i=1}^np^i \sum_{t'\in\mathcal{T}}\left(w(t')r(d_i)\prod_{j=1}^{i-1}(1-r(d_j,t'))\right)-e\sum_{i=1}^np^i+\\
&\sum_{t'\in\mathcal{T}}\left(w(t')r(d_{n+1})(1-r(d_n,t'))\prod_{j=1}^{n-1}(1-r(d_j,t'))\right)-ep^{n+1}\\
\end{align*}
}
Given that $\forall t'\neq t \ldotp r(d_{n+1},t')=0$, it is equivalent to:
{\scriptsize
\begin{align*}
\mbox{RBU}\left(\vec{d},d_{n+1}\right)&=\sum_{i=1}^np^i \sum_{t'\in\mathcal{T}}\left(w(t')r(d_i)\prod_{j=1}^{i-1}(1-r(d_j,t'))\right)-e\sum_{i=1}^np^i+\\
&\left(w(t)r(d_{n+1})(1-r(d_n,t'))\prod_{j=1}^{n-1}(1-r(d_j,t))\right)-ep^{n+1}\\
\end{align*}
}
Given that $1-r(d_n,t')=0$, we obtain:
{\scriptsize
\begin{align*}
\mbox{RBU}\left(\vec{d},d_{n+1}\right)&=\sum_{i=1}^np^i \sum_{t'\in\mathcal{T}}\left(w(t')r(d_i)\prod_{j=1}^{i-1}(1-r(d_j,t'))\right)-e\sum_{i=1}^np^i+0=\mbox{RBU}\left(\vec{d}\right)\\
\end{align*}
}
}
\end{proof}
\vspace{-01.2em}
\vspace{-1.3mm}
\begin{proof}
\textit{\RBU satisfies the \AspRel constraint (Eq.~\ref{eq:AspRel}).}

{\small
 Under the constraint conditions:
{\scriptsize
\begin{align*}
& \mbox{RBU}\left(\vec{d}_{d_i\leftrightarrow d_i'}\right)>RBU\left(\vec{d}\right)\Leftrightarrow\\
& w(t')r(d',t')\prod_{j=1}^{i-1}(1-r(d_j,t'))>w(t)r(d,t)\prod_{j=1}^{i-1}(1-r(d_j,t))\Leftrightarrow \\
& w(t')r(d',t')\prod_{j=1}^{i-1}1>w(t)r(d,t)\prod_{j=1}^{i-1}1\Leftrightarrow w(t')>w(t)
\end{align*}
}
\vspace{-3em}
}
\end{proof}